\documentclass[12pt, preprint]{aastex}

\newcommand{\K}{\,{\rm K}}

\shorttitle{The Inner Rim of YSO Disks}
\shortauthors{Tannirkulam et al.}

\begin{document}




\title{The Inner Rim of YSO Disks: Effects of dust grain evolution.}



\author{A.~Tannirkulam\altaffilmark{1},
T.~J.~Harries\altaffilmark{2}, J.~D.~Monnier\altaffilmark{1}
}
\altaffiltext{1}{atannirk@umich.edu: University of Michigan, Astronomy Dept, 
500 Church Street, 1017 Dennison Bldg, Ann Arbor, MI 48109-1042}
\altaffiltext{2}{University of Exeter, School of Physics, Stocker Road, Exeter, EX4 4QL}


\begin{abstract}
\noindent Dust-grain growth and settling are the first steps towards planet
formation. An understanding of dust physics is therefore integral to a
complete theory of the planet formation process. In this paper, we
explore the possibility of using the dust evaporation front in YSO
disks (`the inner rim') as a probe of the dust physics operating in
circumstellar disks.  The geometry of the rim depends sensitively on the
composition and spatial distribution of dust. Using radiative transfer
and hydrostatic equilibrium calculations we demonstrate that dust
growth and settling can curve the evaporation front dramatically (from a cylindrical radius of about 0.5 AU in the disk mid-plane to 1.2 AU in the disk upper layers for an A0 star). We compute synthetic images and interferometric visibilities for our  representative rim models and show that the current generation of near-IR
long-baseline interferometers (VLTI, CHARA) can strongly constrain the dust properties of circumstellar disks, shedding light on the relatively poorly understood processes of grain growth, settling and turbulent mixing.
\end{abstract}




\section{Introduction}
\label{intro}
Advances in long-baseline near-infrared interferometry have made it
possible to study the inner regions of circumstellar disks at
sub-AU scales.  Early results \citep{rmg1999, rmg2001, lkha2001,
monnier2002a} have shown that the dust disk gets truncated at a finite
radius (determined by the luminosity of the central star and dust
sublimation temperature) within which the temperature is too high for
dust to survive. The truncated disk forms a `rim'
\citep[hereafter DDN01]{natta2001, dullemond2001} that subtends a significant
solid-angle around the star and intercepts stellar photons,
re-radiating predominantly in the near-IR. This rim naturally explains
the near-IR excess observed in Herbig Ae systems.

Despite its success in accounting for the near-IR excess, the vertical
rim model introduced in DDN01 has had a few weaknesses from the very
outset. The flux received from a vertical rim is highly viewing-angle
dependent, since the projected line-of-sight area of the rim changes
rapidly with inclination angle. This implies that the near-IR spectral
energy distributions (SEDs) of
Herbig Ae/Be stars should be a strong function of inclination
angle, but observations show that most Herbig Ae stars have
similar near-IR excesses, independent of their inferred inclination
angles \citep{natta2001, dominik2003}. Furthermore, the surface brightness
distribution of a vertical rim will become highly asymmetric at
inclination angles different from face-on. This asymmetry should be
detectable by interferometers in the form of a closure-phase signal
\citep[and references therein]{monnier2000}, but  recent studies
\citep{monnier2006} do not detect significant closure-phase signals
indicating that the geometry of the inner rim is not a vertical wall.

It appears that a mechanism that curves the inner rim is needed.
\citealt*{isella2005} (2005) (hereafter IN05) studied the impact of
density-dependent dust sublimation temperatures on the structure of
the rim. Close to the mid-plane, where the densities are high, the dust
sublimation temperatures are also high \citep{pollack}, while as the
densities begin to fall rapidly in the vertical direction the
sublimation temperatures drop. The IN05 calculations show that this
causes the inner rim to curve away from the surface of the star. \citet{isella2006} found that a curved-rim model provides a substantially better fit to the
visibilities and near-IR SEDs of a few Herbig Ae systems.

In this paper we look at another mechanism that could curve the rim on
larger scales than the ones produced by a density-dependent
dust sublimation model---namely, dust growth and settling. In the
absence of strong turbulence, dust grains will settle towards the disk
mid-plane with large grains settling lower and having a smaller scale
height than small grains. Dust grains will also grow rapidly in the
mid-plane due to the much higher densities \citep{dubrulle1995,
dullemond2004, tanaka2005}. The opacity in the mid-plane is thus
dominated by large grains whereas small grains dominate the opacity
higher in the disk photosphere. This transition in the nature of the
opacity source along with dust sublimation curves the inner rim as
described in Section~4. Near-IR interferometers with baselines several
hundred meters long (e.g. CHARA, VLTI) can distinguish between the model
presented in this paper and the IN05 model, shedding light on dust
physics operating at sub-AU scales.

We have computed the structure of the dust evaporation front in a
self-consistent manner for the first time in a 2-D geometry \citep[photon
scattering has not been included; for effects of scattering on the disk structure and SED see][]{dullemond2003} using the  radiative
transfer code {\sc torus} \citep{harries2000}. In the following section we
describe the radiative transfer code, while in section~3 we perform a
consistency check by comparing the results for a {\sc torus} generated
density-dependent dust sublimation model with IN05 results. In Section~4
we present results for the dust segregation model, followed by a summary
of the results and conclusions in Section~5.

\section{The Monte Carlo Radiative Transfer Code - TORUS}
\subsection{Description of grid and disk structure calculations}

The calculations in this paper were performed using the {\sc torus}
Monte-Carlo radiative-transfer code \citep{harries2000, harries2004,
kurosawa2004}. Radiative equilibrium is computed using Lucy's
\citep{lucy1999} algorithm on a two-dimensional, cylindrical-cartesian
adaptive-mesh grid. Storing the opacity and temperature information on
an adaptive mesh has particular advantages for the problem considered
here, since the accurate determination of temperatures and subsequent
SEDs requires an adequate resolution of the effective disk
photosphere, whose position is changing spatially both as the dust is
sublimated and as the vertical structure of the disk changes as it is
iterated towards hydrostatic equilibrium. Using adaptive-mesh
refinement (AMR) we are able to subdivide cells near the disk
photosphere as the calculation proceeds, ensuring the resolution is
maintained at each step.

The initial density structure for the disk calculations is based on
the canonical description of the $\alpha$-disk developed by
\citet{shakura1973}, viz
\begin{equation}
\rho(r,z) = \rho_0 \left (\frac{r}{r_0}\right)^{-\alpha}\exp\left[ -\frac{1}{2}\frac{z^2}{h(r)^2}\right]
\end{equation}
where $r$ is the radial distance in the mid-plane, $r_0$ is some characteristic radius, $z$ is the distance perpendicular to the mid-plane, 
and $h(r)$ is the scaleheight, given by
\begin{equation}
h(r) = h_0 \left(\frac{r}{r_0}\right)^{\beta}
\end{equation}
with parameters of $\alpha=2.625$ and $\beta=1.125$, giving a radial
density dependence of the surface density of $\Sigma(r) \propto
r^{-1.5}$. The AMR mesh is divided such that the cells are
logarithmically spaced in the radial direction and that there are at
least 7 cells in the vertical direction per disk scale height
$h(r)$. A sweep of the grid is then made to split cells further around
the $\tau_{5500}=1$ ($\tau_{5500}$ is optical depth calculated at 5500 \AA) surface using the following algorithm: for each
pair of adjacent AMR cells the individual optical depths across the
cells are calculated ($\tau_1$ and $\tau_2$ say, with $\tau_1$
referring to the more optically thick of the two). The more optically
thick cell is subdivided providing that $\tau_2 < \tau_{\rm min}$ and
$\tau_1 > \tau_{\rm max}$, where $\tau_{\rm min}$ and $\tau_{\rm max}$
are parameters (typically 0.1 and 0.5 respectively). Should any cell
in the grid be subdivided, the sweep is repeated until no new cells
are added to the mesh.

Once the temperature (we assume that the disk is in local
thermodynamic equilibrium and gas and dust are thermally coupled) and
dust sublimation (see next section) structures have converged using
the Lucy algorithm, the vertical disk structure is modified via the
equation of hydrostatic equilibrium following a similar algorithm to
that detailed by \citet{walker2004}. Briefly, and assuming disk mass
is negligible compared to the central star, the equation of
hydrostatic equilibrium is
\begin{equation}
\frac{{\rm d}P}{{\rm d}z} = -\rho g_z
\end{equation}
where $P$ is the pressure and $g_z$ is the vertical component of the
stellar gravity. Adopting an ideal gas equation of state $P=\rho
kT/\mu m_{\rm H}$ (with $\mu=2.3$ for a standard molecular
hydrogen-helium mixture) we get
\begin{equation}
\frac{{\rm d}\rho}{{\rm d}z} \frac{1}{\rho} = -\frac{1}{T} \left(
\frac{g_z \mu m_{\rm H}}{k} + \frac{{\rm d}T}{{\rm d}z} \right).
\end{equation}
The above equation can be solved for $\rho(z)$ numerically on the AMR
mesh since the vertical temperature structure is known from the
radiative-equilibrium calculation. The vertical density structure is
then renormalized to conserve the radial surface density profile
detailed above. A self-consistent calculation for dust sublimation (see \S\ref{methodology}) and disk temperature followed by a hydrostatic
equilibrium calculation is repeated until the disk density
structure has converged. Convergence is typically achieved in four
iterations. Images and SEDs are subsequently calculated using a
separate Monte Carlo algorithm based on the dust emissivities and
opacities \citep{harries2000}.

\label{hydro}
\subsection{Implementation of Dust Sublimation.}
\label{methodology}
In the models discussed here the shape of the inner rim is controlled
by dust sublimation. Typical mid-plane densities in the dust sublimation region are of the order of $10^{-8}-10^{-10}$ g\,cm$^{-3}$ in Herbig disks. In the mid-plane, an optical depth of unity for visible light photons (measured radially inward into the disk from the rim edge) corresponds to a length scale of $\sim$ $10^{-5}-10^{-7}$ AU. Therefore if we started sublimation runs with a
grid that had the full dust opacity present, convergence towards
equilibrium rim shape would be extremely slow. This is very similar to
the problem encountered in photoionization codes, where neutral
hydrogen opacity to ionizing photons is so large that starting from a
neutral grid would take an impractical amount of CPU time for the
Str\"omgren sphere to propagate outwards \citep{wood2004}.

The solution that we adopt is also analogous to the one used in
photoionization codes \citep{wood2004}. The dust content is first reduced to a very low value in the computational grid for the circumstellar disk, to make each of the grid cells optically thin. Stellar photons then propagate through the disk
and the temperature of grid cells is determined. Dust is added to
cells that are cooler than the sublimation temperature in small steps of $\tau$.
 The step size is a $\tau$ of $10^{-3}$ (computed at 5500\AA) for the first five dust growth steps. The step size is then increased
logarithmically, first to $10^{-2}$, then to $10^{-1}$ and so on until
the appropriate gas to dust ratio is reached in each grid cell.  The
grid cell temperatures are recomputed after every dust growth step and the process is repeated until the shape of the dust sublimation region converges.

\begin{figure}[thb]
\begin{center}
{
\includegraphics[angle=90,width=5in]{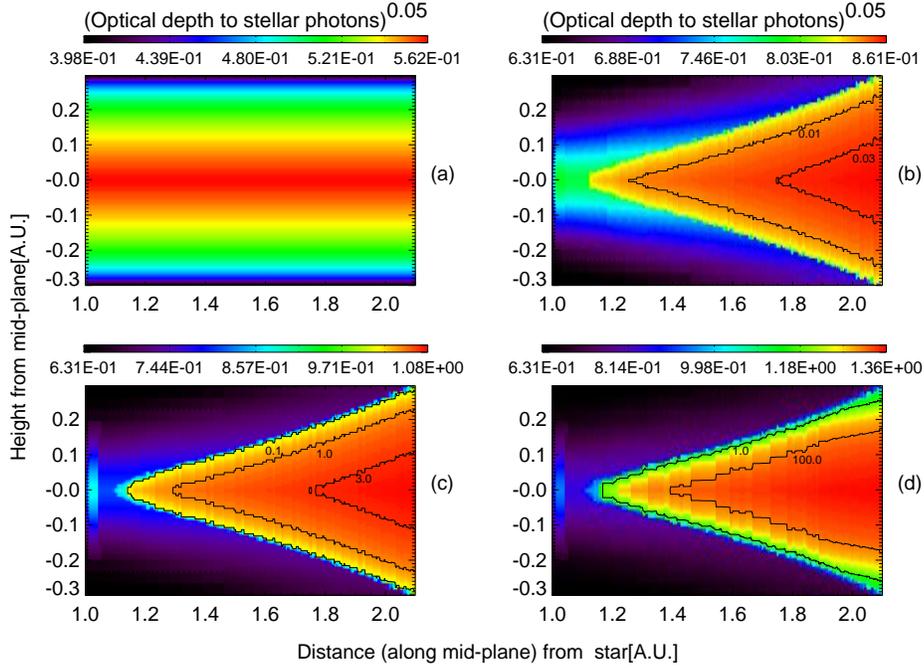}
}

\caption{a) Cross section of an inner rim from which dust
has been stripped. Photons are propagated through this rim to
determine the `optically thin' grid-cell temperatures. The color
scheme shows integrated $\tau$, measured along lines perpendicular to
the disk mid-plane. b) Inner rim after the first dust growth
step. Dust is grown in cells that are cooler than the sublimation
temperature. The contours connect points with equal integrated tau. c)
and d) depict stages further along in the dust growth scheme. The geometry
of the rim is ``sublimation converged'' in (d) (see
\S\ref{methodology}). A hydrostatic equilibrium calculation (see
\S\ref{hydro}) is then performed. The dust growth and hydrostatic
equilibrium calculations are repeated until convergence is reached for
the structure. \it {The rim shapes depicted above are not the final converged solution. See Figure \ref{comparisona} for the final rim shapes.}
\label{figmethodology}}
\end{center}
\end{figure}

Figure~\ref{figmethodology} illustrates the dust-growth process in our
computational scheme. The color scheme shows integrated optical depth, measured
along lines perpendicular to the disk mid-plane. Panel-(a) is a grid
for which the optical depth across each cell is $10^{-8}$. Panel-(b) is a cross section of the grid after the first
dust growth step where $\tau$  across each cell is now $10^{-3}$. The cells have a $\tau$ of $10^{-1}$
across them in panel-(c). The optical depth across each of the cells that lie away from the rim surface is 10 in panel-(d), whereas cells at the outer surface of the rim have been
smoothed to resolve the $\tau=1$ surface. The geometry of the rim can considered to be `sublimation converged' in panel-(d). Any further addition of dust does not significantly change the distance  of the $\tau=1$ rim surface from the star.  An appropriate amount of dust is put into each of the cells to reach a predetermined gas to dust
ratio in the final dust growth steps.  A hydrostatic equilibrium
calculation as described in ~\S\ref{hydro} is then performed. The dust
growth and hydrostatic equilibrium calculations are repeated until
convergence for the entire disk structure is reached (typically 4
iterations). In the next section we describe checks that ensure that
this process gives accurate results.
\section{Code testing against IN05 results}
\label{IN05res}
IN05 calculated inner rim models with astronomical silicate dust
 \citep{weingartner2001}, where the shape of the rim is set by the
 fact that the evaporation temperature of dust depends on gas density
 \citep{pollack}. In this model, silicate grains sublimate at a higher
 temperature compared to other grains and hence fix the rim location.
 For silicate dust, the evaporation temperature $T_{\rm evp}$ can be parameterized as
 \begin{equation}T_{\rm evp} = G\left[\frac{\rho_{\rm gas}(r,z)}{\mbox{1g }\mbox{cm}^{-3}}\right]^\gamma\end{equation}
 where $G=2000\K$, $\gamma = 1.95\times10^{-2}$ and $\rho_{\rm gas}$ is the
 density of gas in g\,cm$^{-3}$ (see IN05 eq. [16]). IN05 showed that this
 dependence of $T_{\rm evp}$ on gas density curves the inner rim.

\begin{figure}[thb]
\begin{center}
\includegraphics[angle=90,height=2in, width=7in]{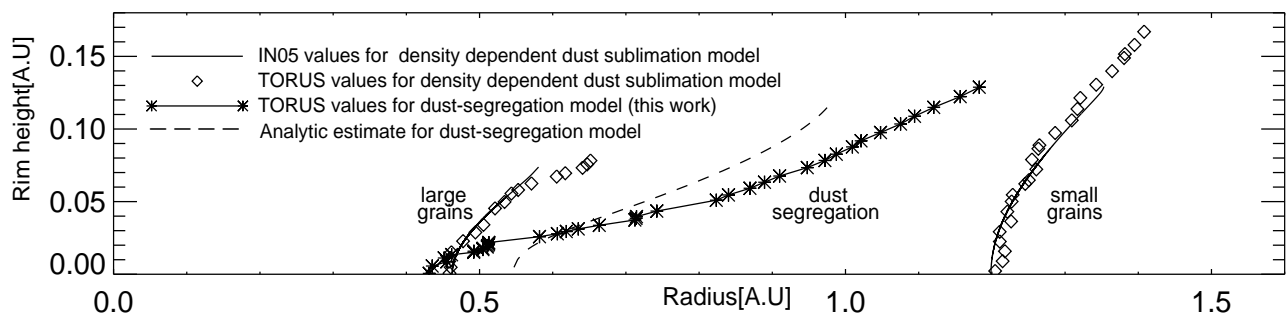}
\caption{The `rim' is defined as the $\tau =1$ surface
(for $\lambda$= 5500\AA), computed along radial lines from the central
star. The figure shows the height of the inner rim above the disk
mid-plane. The IN05 rim has been scaled at the $\sim8\%$ level to
match up with the {\sc torus} rim. The dashed lines are an analytical
estimate of the evaporation front for the dust segregation model (see
\S\ref{analytics}).
\label{comparisona}}
\end{center}
\end{figure}

\begin{figure}[thb]
\begin{center}
\includegraphics[angle=90,height=4in]{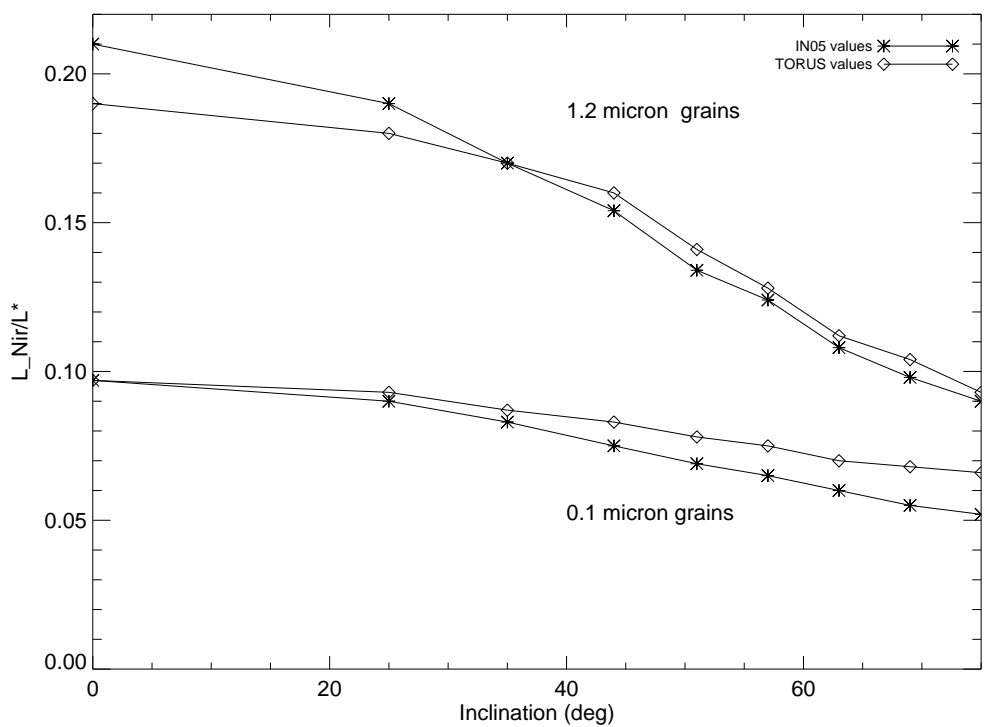}
\caption{Near infrared emission (integrated between 1.25--7$\mu$m) from the inner rim as a function of inclination angle,
plotted for small and large grains. The emission has been normalized
to the stellar luminosity.
\label{comparisonb}}
\end{center}
\end{figure}

In order to test the numerical scheme described in \S\ref{methodology} we compared
{\sc torus} results for the rim with the IN05 results (kindly provided by A. Isella). Table~\ref{table1}
describes the properties of the star and the circumstellar disk used
in the comparisons. Figure~\ref{comparisona} shows the shape of the
rim for large (1.2$\mu$m) and small (0.1$\mu$m) grains. The rim here
is defined as the $\tau =1$ surface (for $\lambda$ = 5500\AA),
calculated along radial lines from the central star towards the
disk. Figure~\ref{comparisonb} compares the
fraction of stellar luminosity re-emitted by the rim in infrared
wavelengths (integrated over 1.25--7$\mu$m) as computed by IN05 and
{\sc torus}.

Figures~\ref{comparisona} and \ref{comparisonb} show that that the
{\sc torus} results are consistent with the IN05 calculations at
better than 15\% level, giving us confidence in our Monte Carlo (MC)
computations. We note that benchmark comparisons for MC codes,
studying radiative transfer in optically thick disks (Pascucci et
al. 2004), show a similar level of agreement among the various MC
codes.

\section{Rim curvature due to grain growth/settling}

\subsection{Model Description}
\label{model1}
The rim in the IN05 model is curved because of a density-dependent
dust-evaporation temperature. In this section we describe an alternative
model in which the  dust sublimation temperature is kept fixed at 1400\,K and rim curvature arises entirely because of dust growth and settling.

Small grains are much less efficient at cooling than large grains. If
only single-sized grains are present, then at a given distance from
the star, small grains will have a higher temperature than large
grains. Under the assumption of a good temperature coupling between
gas and dust \citep[at densities characteristic of the rim region,
this is a good approximation even in the disk
photosphere;][]{kamp2004}, a mixture of small and large grains will be
hotter than the case where only large grains are present. Combined
with dust sublimation, this means that the location and shape of the
inner rim will change depending on the relative fraction of small and
large grains.

A number of observational \citep{alessio2001, alessio2006, chiang2001, rettig2006}
and theoretical \citep{dullemond2004, tanaka2005} results have shown
that dust properties in circumstellar disks evolve with time. Small
grains tend to coagulate into larger grains and settle towards the
disk mid-plane, with the evolution of dust grain sizes occurring most
rapidly in the disk mid-plane where the densities are high. In the
absence of strong turbulence (the nature and strength of turbulence in
circumstellar disks is not well understood), large grains will have a
smaller scale-height than small grains.  Thus grain growth and settling
provide a natural means for changing the fraction of large and small
grains as a function of height (`dust segregation') from the disk
mid-plane. This will curve the rim as described in the previous paragraph.

As an illustration, we look at a model (henceforth `dust segregation
model') with silicate dust \citep{weingartner2001}. In order to
simulate dust settling  we fix the large (1.2$\mu$m) grain
scale height to be 60\% of the gas scale height. We do not settle  the small grains (0.1$\mu$m) and their scale height is fixed to be the same as the gas scale height. As grains grow, the net mass of small grains decreases, while the net mass of
large grains increases: we fix the mass of large
grains to be $\sim$ 9 times the mass of small grains. All the the dust mass resides in the 1.2 \& 0.1 $\mu$m grain components and  total mass of dust in the disk is chosen to be 1/100th of the total
gas mass. The rim gets curved over a wide range of parameter choices. The curvature can be made stronger by increasing the relative fraction of small grains.

The above description leads to the following expressions for dust density:
\begin{equation}
\rho_{1.2} = 1.5\times 10^{-2}\rho _{o\_{\rm gas}}\exp \left[ -\frac{1}{2}\frac{z^2}{(0.6h)^2}\right]
\end{equation}

\begin{equation} 
\rho_{0.1} = 10^{-3}\rho_{o\_{\rm gas}} \exp \left[ -\frac{1}{2}\frac{z^2}{h^2}\right]
\end{equation}
where $\rho_{1.2}$ is the density of 1.2$\mu$m grains, $\rho_{0.1}$
is the density of 0.1$\mu$m grains, $\rho_{o\_{\rm gas}}$ is the gas
density in the mid-plane, $z$ is distance perpendicular to the disk mid-plane and $h$ is the gas scale height. The star and disk parameters are
chosen as in Table~\ref{table1} (the parameters are the same as in the
IN05 models of section 3).

\subsection{Geometry of the Dust Segregation Rim} 

\subsubsection{Rim Shape}
\label{shape}
Figure~\ref{comparisona} shows the shape of the inner rim for the dust
segregation model. The curvature of the rim depends on the relative
density fractions of large and small grains and the width depends on
the sizes of the two dust components.  A major finding here is that
the dust segregation model can curve the rim on much larger scales than a
density-dependent dust sublimation temperature (IN05) model. The $\tau =
1$ surface is nearly vertical at the inner edge of the dust
segregation model because of the fixed, density-independent dust
sublimation temperature.

\subsubsection{Analytic Estimate for Rim Shape} 
\label{analytics}

The most important dust parameter that determines the rim shape is
$\epsilon(z)$, defined as the ratio of Planck mean opacities at the
dust sublimation temperature and the stellar effective temperature
(IN05). Using equations (6) and (7) and values of Planck mean opacities for
0.1$\mu$m and 1.2$\mu$m grains, $\epsilon$ (see appendix for a more general expression)
for the dust segregation model is given by

\begin{equation}
\epsilon\left (z\right) = \left[ \frac {0.42 + 0.015 \exp
  \left(\frac{0.89z^2}{h^2}\right)}{0.76 + 0.181 \exp \left(\frac{0.89z^2}{h^2}\right)} \right]
\end{equation}

\begin{figure}[thb]
\begin{center}
{
\includegraphics[angle=90,width=3in]{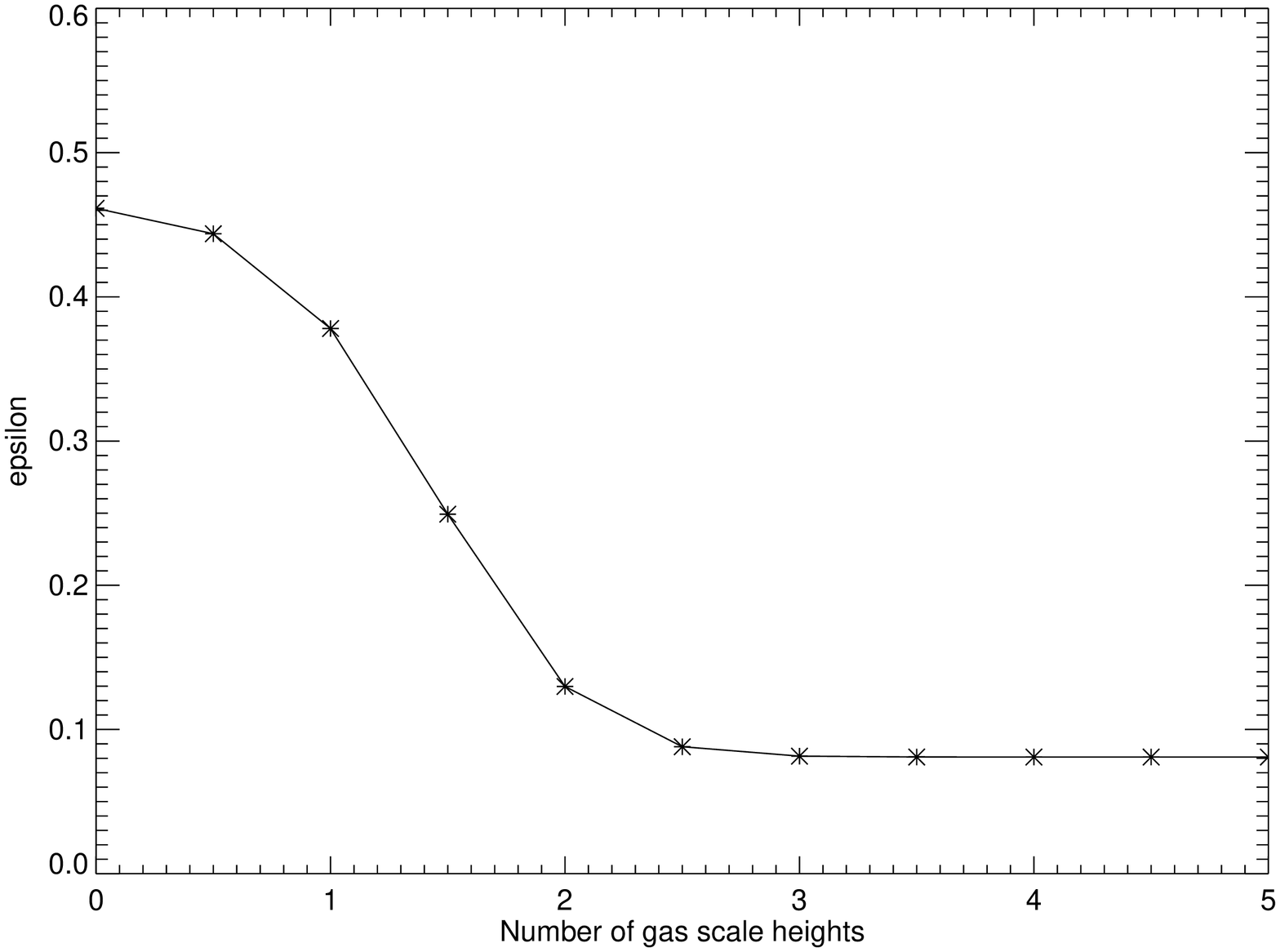}
\hphantom{.....}
\includegraphics[angle=90,width=3in]{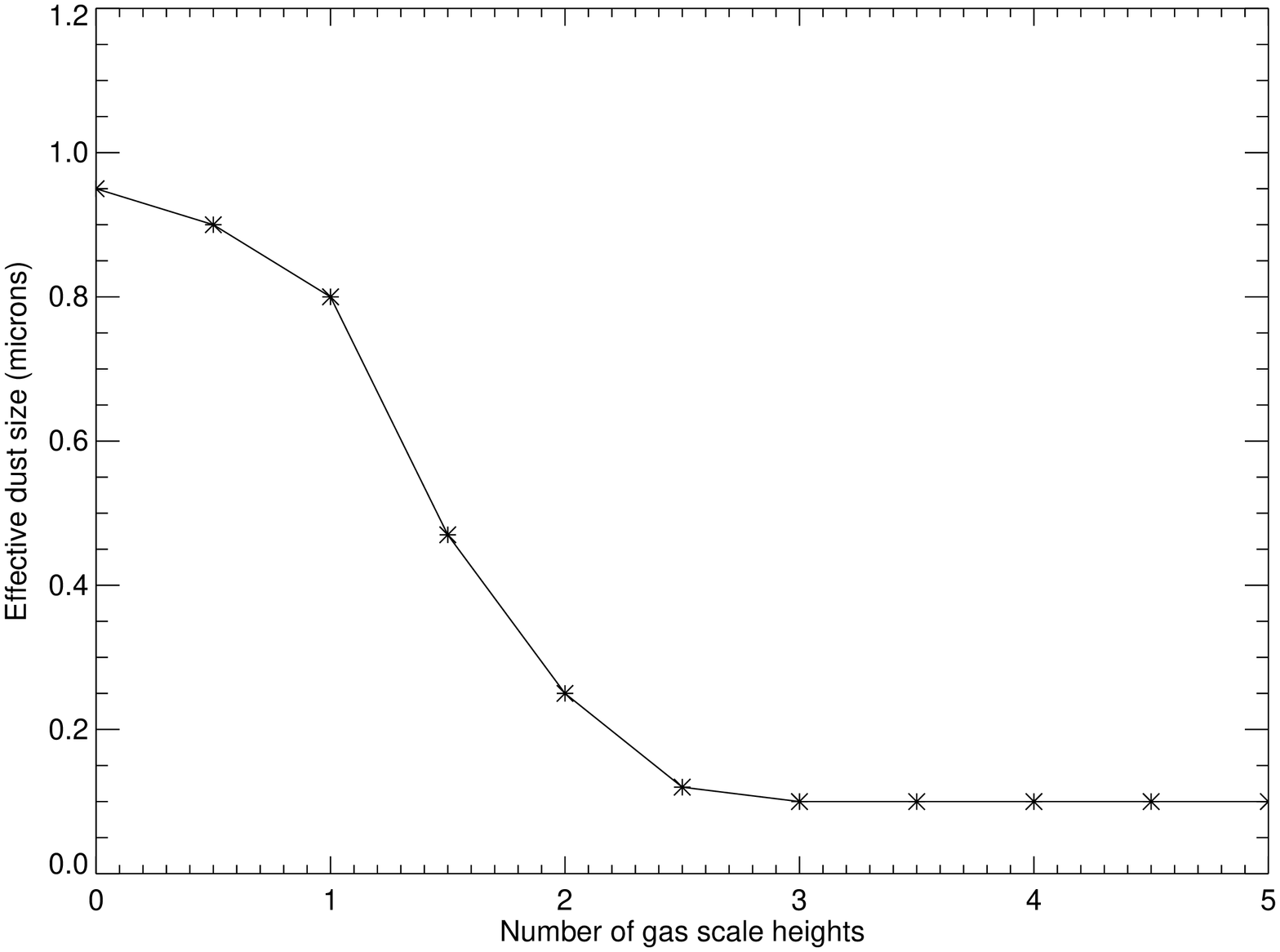}
}

\caption{a) The ratio of Planck mean opacities for dust at
the sublimation temperature and at the stellar photospheric
temperature for the dust segregation model.  b) Effective dust size
plotted against the number of gas scale heights (see \S\ref{model1}).
\label{dust_size}}
\end{center}
\end{figure}

\noindent Figure~\ref{dust_size}a shows $\epsilon$ as a function of mid-plane
 height. Corresponding to this $\epsilon$, an effective dust size
 which has the same $\epsilon$ as the dust mixture, is plotted in
 Figure~\ref{dust_size}b.

A crude estimate for the rim shape in the dust segregation model can
be obtained analytically. The dust destruction radius $R_{\rm evp}$
\citep[see][]{calvet1991, isella2006} is

\begin{equation}
R_{\rm evp}\left[\mbox{AU}\right] =  0.034\left[\mbox{AU}\right]\left(\frac{1500\K}{T_{\rm evp}}\right)^2\left[{ \left(\frac{L_*}{L_{\odot}}\right)
\left(B + \frac{1}{\epsilon(z)}\right) }\right]^{1/2},
\end{equation}
$T_{evp}$ is the dust evaporation temperature, $L_*$ is luminosity of the star in solar units and $B$ is a dimensionless diffuse heating term. $B$ characterizes the optical thickness of the rim to its own thermal emission.

The gas scale height $h$ at $R_{evp}$ is estimated to be \citep[see eqs. {[7]}
  and  {[8]} from][]{cg97}
\begin{equation}
 h\left[\mbox{AU}\right] =  R_{\rm evp}\left[\mbox{AU}\right] \left[{\left(\frac{T_{\rm
 gas}}{4\times 10^{7}\K}\right)\left(\frac{M_{\odot}}{M_*}\right)\left(\frac{R_{\rm evp}}{4.64\times 10^{-3} \mbox{ AU}}\right)}\right]^{1/2}
\end{equation}
where $T_{\rm gas}$ is the temperature of the bulk of the gas (assumed to be isothermal) and $M_*$ is mass of the 
star in solar masses.

Combining equations (8), (9), and (10), substituting for stellar parameters
from Table~\ref{table1} and setting $T_{\rm evp}=1400$\,K, the height of
the evaporation front (see appendix for a general result) as a
function of distance from the star along the disk mid-plane becomes
\begin{equation}
 z\left[\mbox{AU}\right] =  1.06R_{\rm evp}\left[\mbox{AU}\right]\left[{\ln\left(\frac{5.87\mbox{AU}^{-2}R^2_{\rm evp}-0.42B-0.76}{0.181 + 0.015B - 0.21\mbox{AU}^{-2}R^2_{\rm evp}}\right)}\right]^{1/2}  \left[{\left(\frac{T_{\rm
       gas}}{1\times 10^{8}\K}\right)\left(\frac{R_{\rm evp}}{4.64\times 10^{-3}\mbox{ AU}}\right)}\right]^{1/2}
\end{equation}

Figure~\ref{comparisona} shows an analytic estimate for the rim shape
\citep[we assume a gas temperature of  1000\,K and have set $B=2$,][]{calvet1991}. The analytic rim agrees with the {\sc torus} dust
segregation rim at the 30\% level. The match at the inner edge of the
rim can be improved by tuning the diffuse heating  parameter $B$, but the match at the outer edge
remains poor because the analytic estimate is not self consistent and
does not take into account the transition between the rim and the
`shadow' (see section \S\ref{images}) region beyond it.

\subsection{Observables}

\subsubsection{SED}
\begin{figure}[thb]
\begin{center}
\includegraphics[angle=90,width=3in]{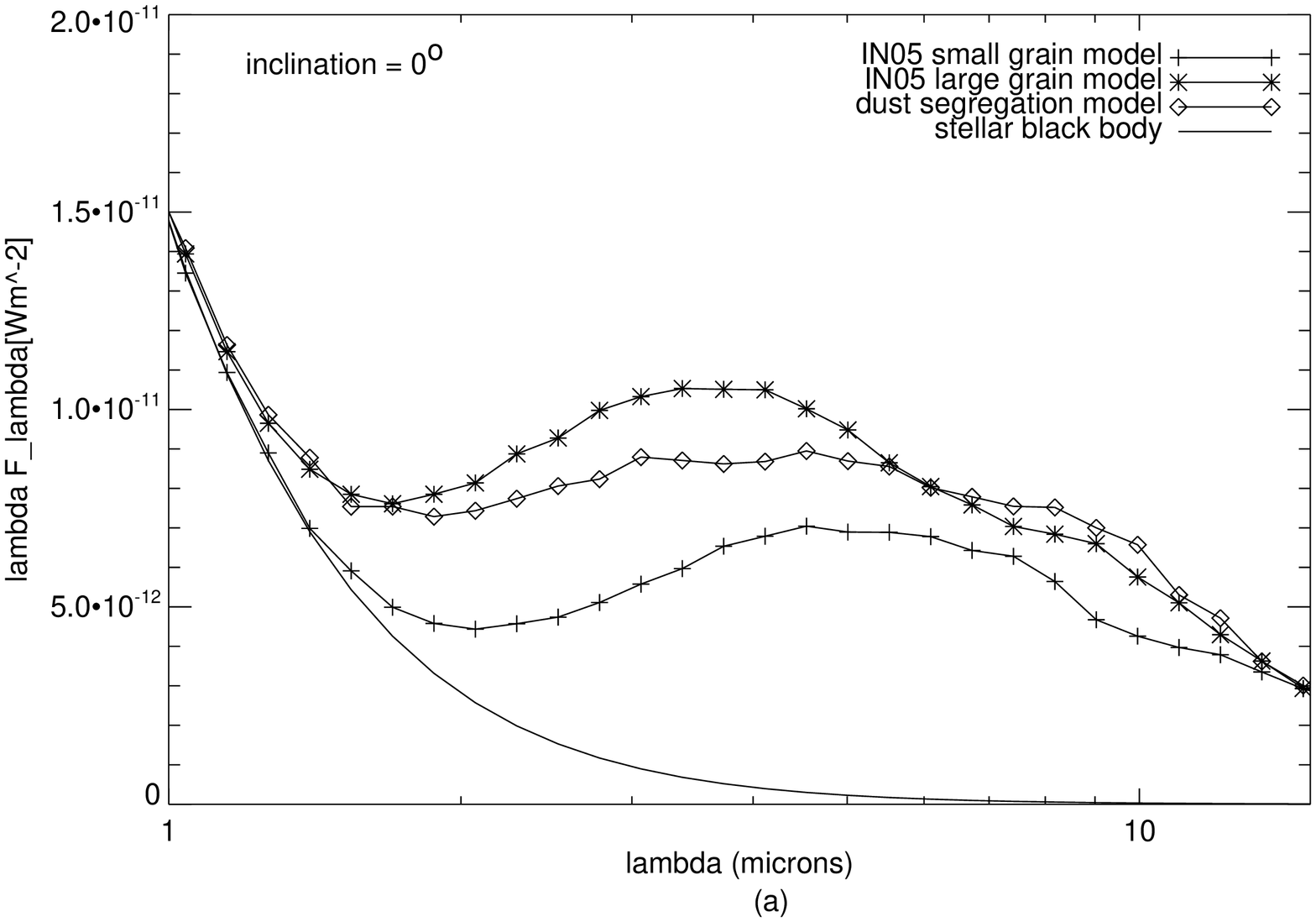}
\hphantom{.....}
\includegraphics[angle=90,width=3in]{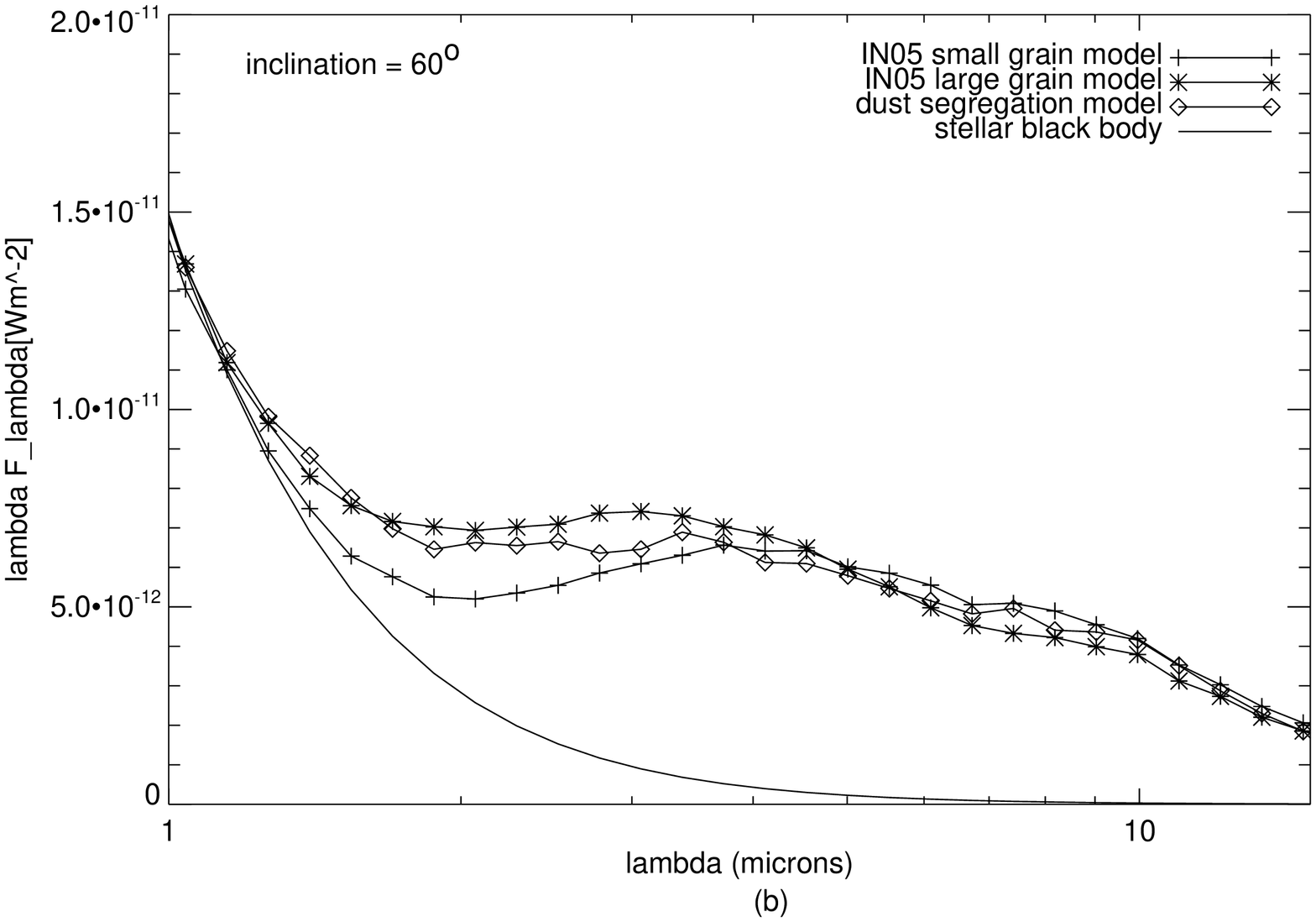}
\includegraphics[angle=90,width=3in]{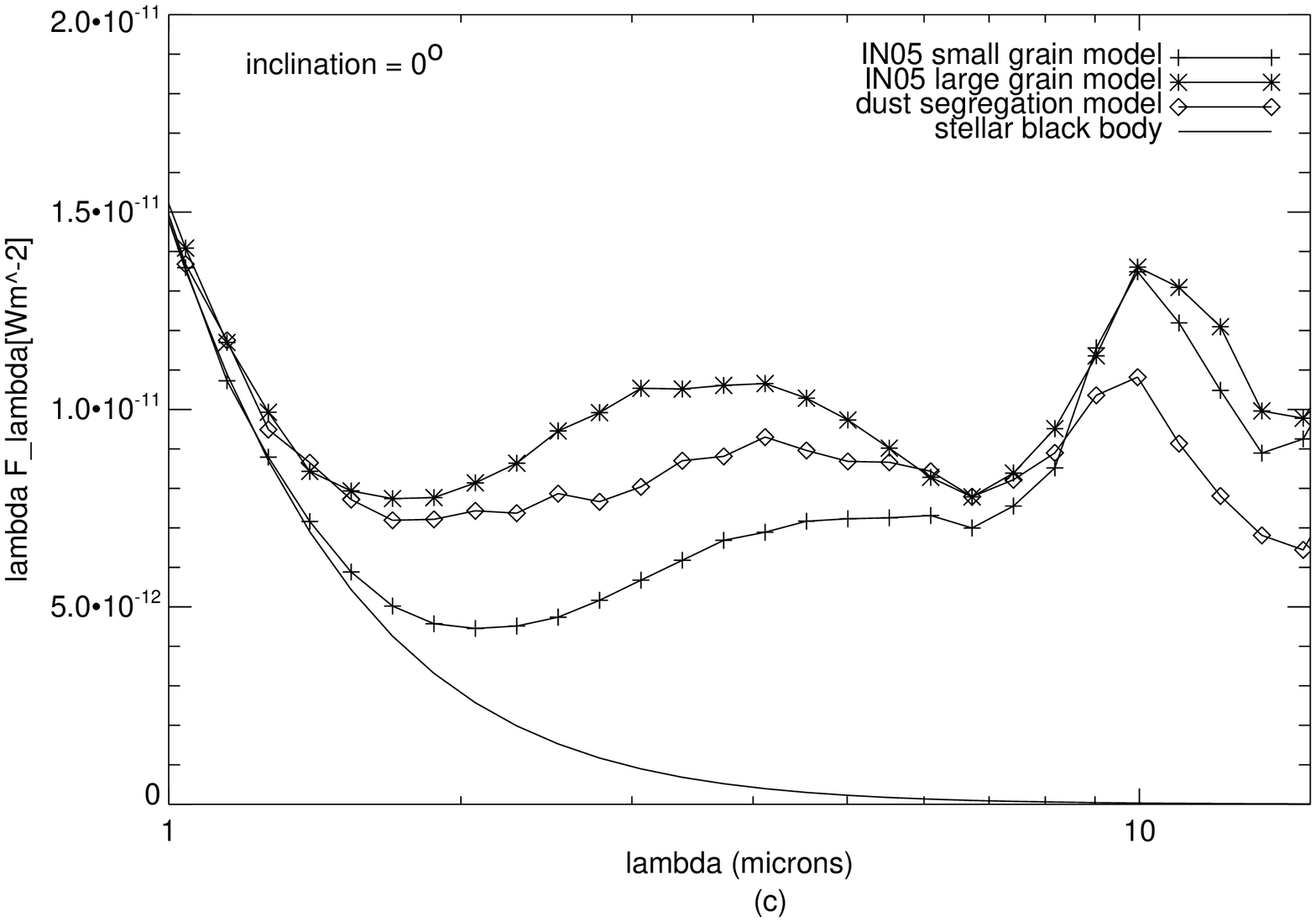}
\hphantom{.....}
\includegraphics[angle=90,width=3in]{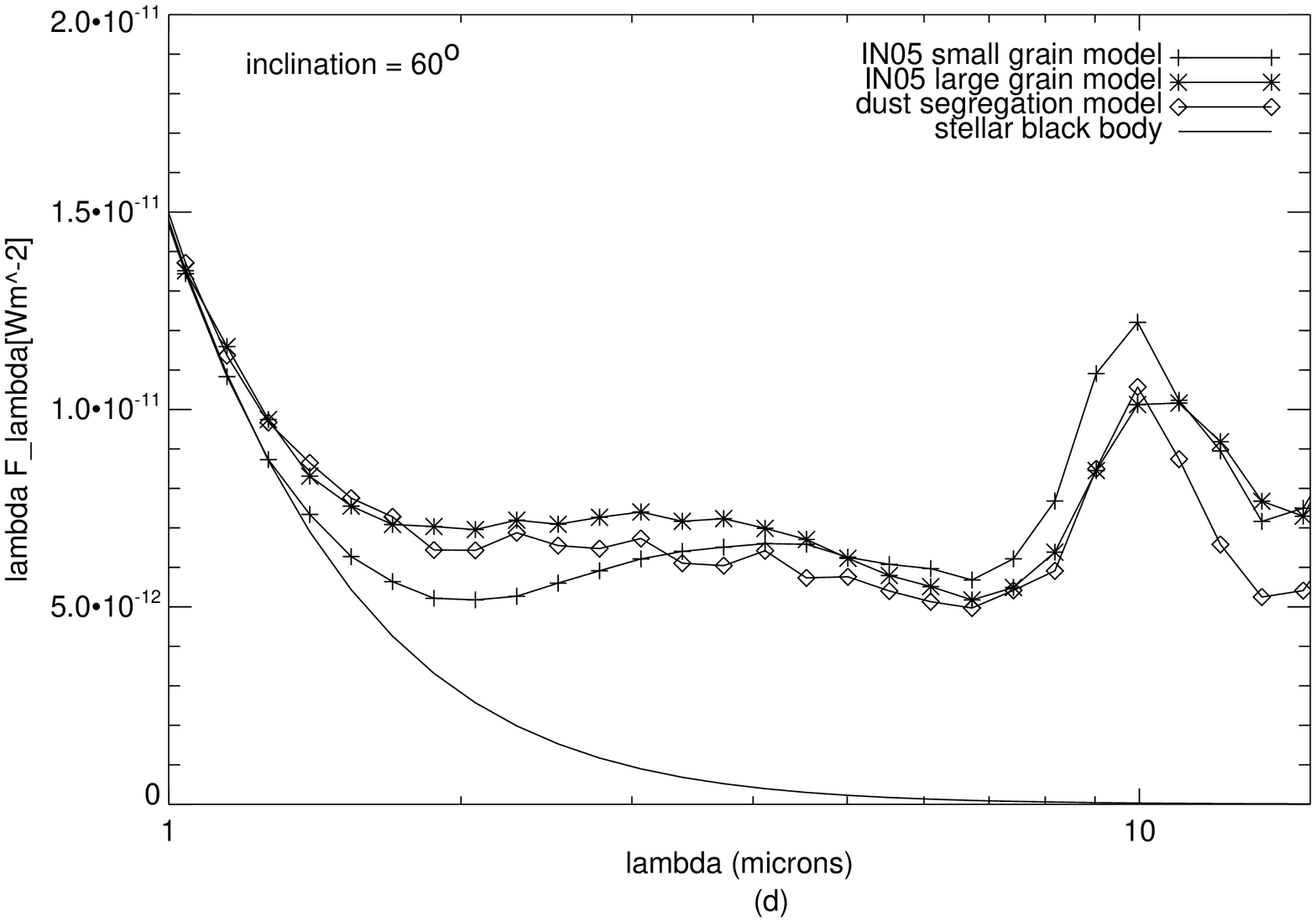}
\caption{a \& b show near and mid-IR SEDs of the star + rim system for the IN05 and dust segregation models. c \& d show SEDs of the star + rim + the disk. The star is placed at 150pc with stellar
parameters described in Table~\ref{table1}. (a, c) system is face-on, (b, d) system is inclined $60^{\mbox{o}}$ from face-on.
\label{dustsegir}}

\end{center}
\end{figure}

Figure~\ref{dustsegir} shows the infrared emission from the inner rim
and the outer flared-disk component. There is significant emission
over the stellar blackbody (denoted by the continuous line) from 1.5$\mu$m onwards. The strength of the rim emission is approximately determined by the temperatures along the rim at optical depths of 2/3 (calculated at the wavelength of emission). Figure 3 in the IN05 paper shows that at a  given optical depth, the small grain rim has a lower temperature than the large grain rim. Because of this temperature difference, near-IR emission in the small grain model gains strength at longer wavelengths compared to the large-grain and dust-segregation models.

\subsubsection{Images and Visibilities}
\label{images}

\begin{figure}[thb]
\begin{center}
\includegraphics[angle=90,height=4.55in, width=7.0in]{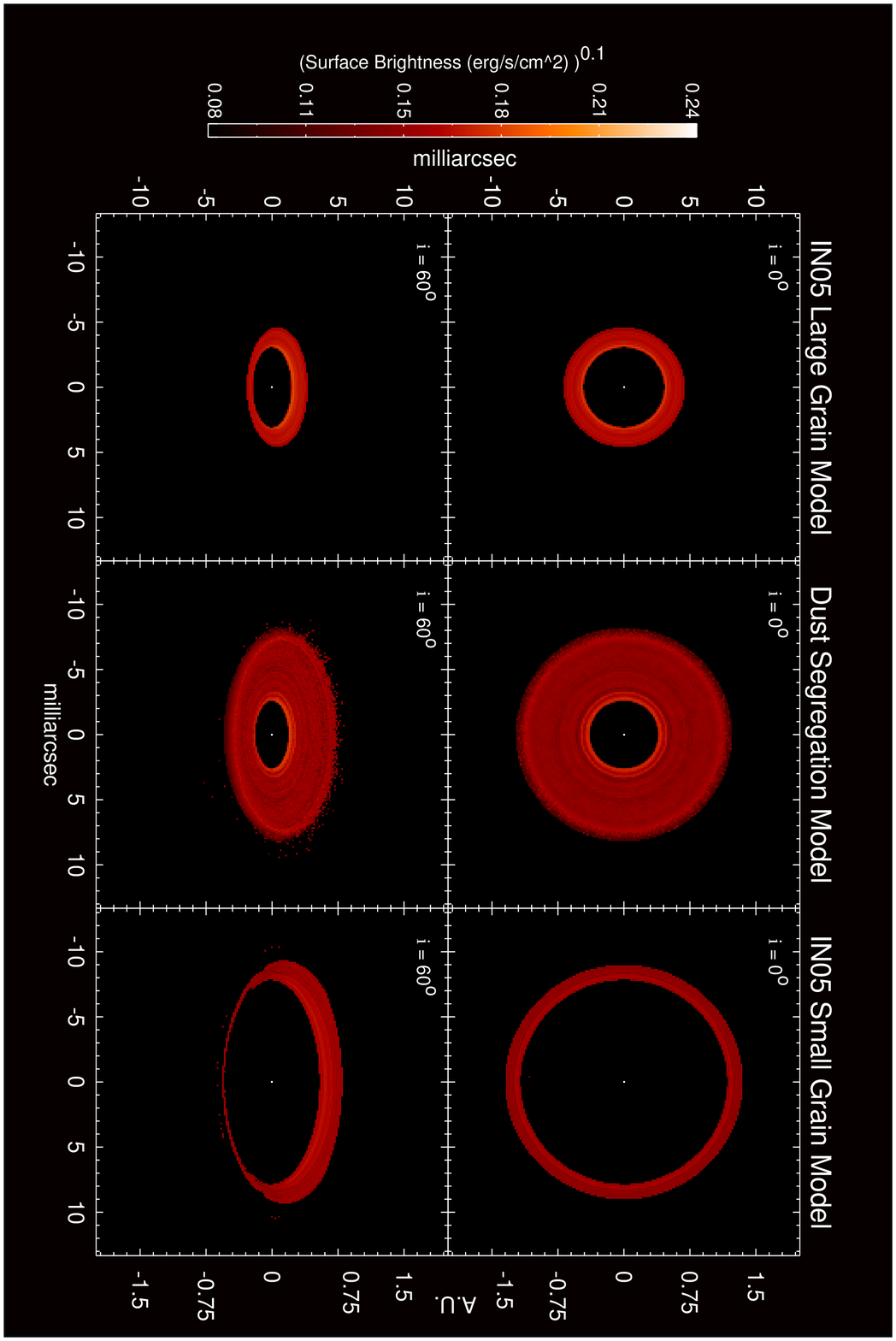}
\caption{Synthetic 2.2$\mu$m images for the different rim
models discussed in the text. The panels on the left and right are IN05
rims computed for 1.2$\mu$m (large grain) and 0.1$\mu$m (small grain) silicate
dust. The center panels are images for the dust
segregation model. The star is placed at 150pc with the stellar
parameters described in Table~\ref{table1}. The star is unresolved
at the image scale and is just one bright pixel at the center of the
images.
\label{dustsegim_2.2}}
\end{center}
\end{figure}

\begin{figure}[thb]
\begin{center}
\includegraphics[angle=90,width=3in]{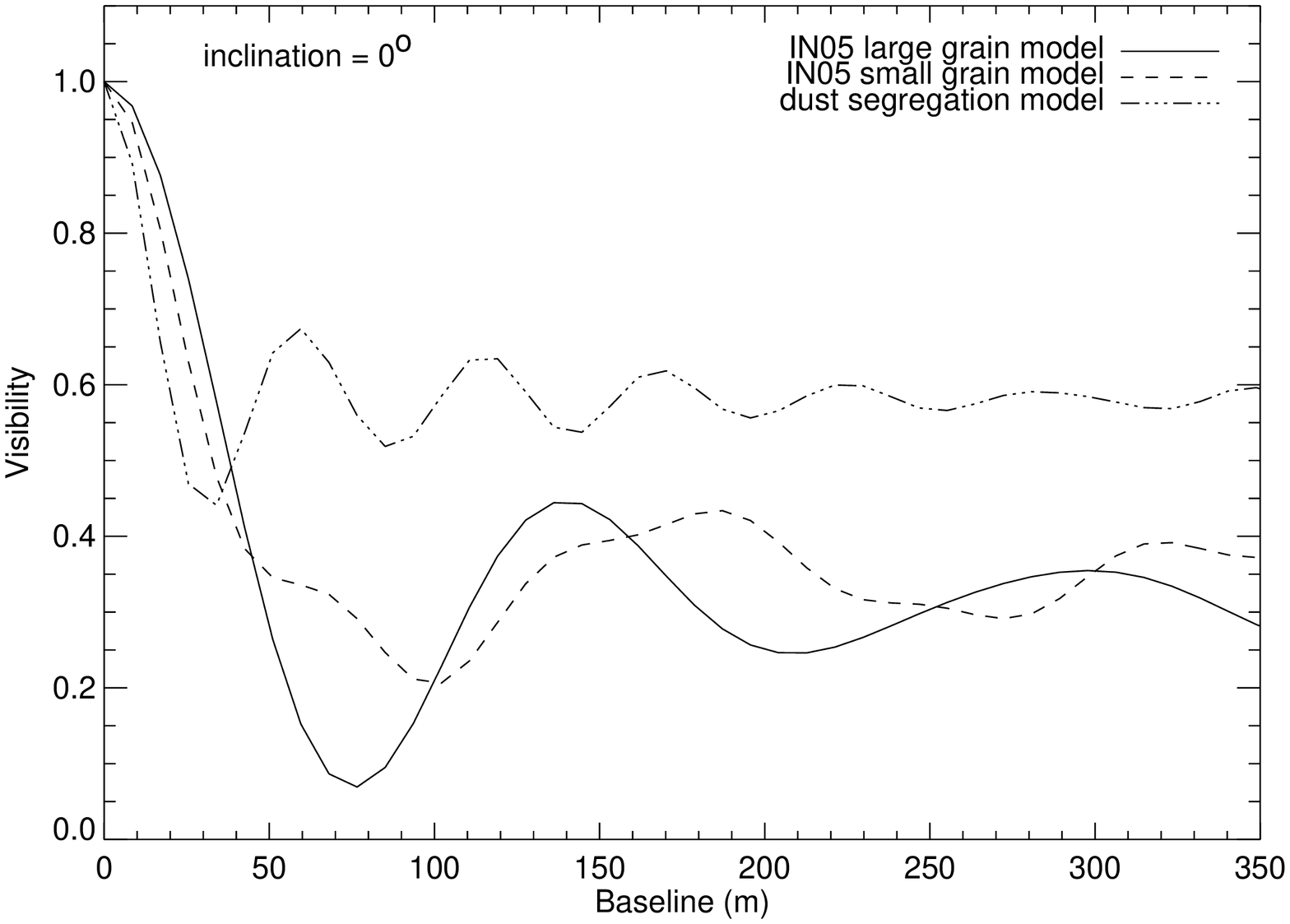}
\hphantom{.....}
\includegraphics[angle=90,width=3in]{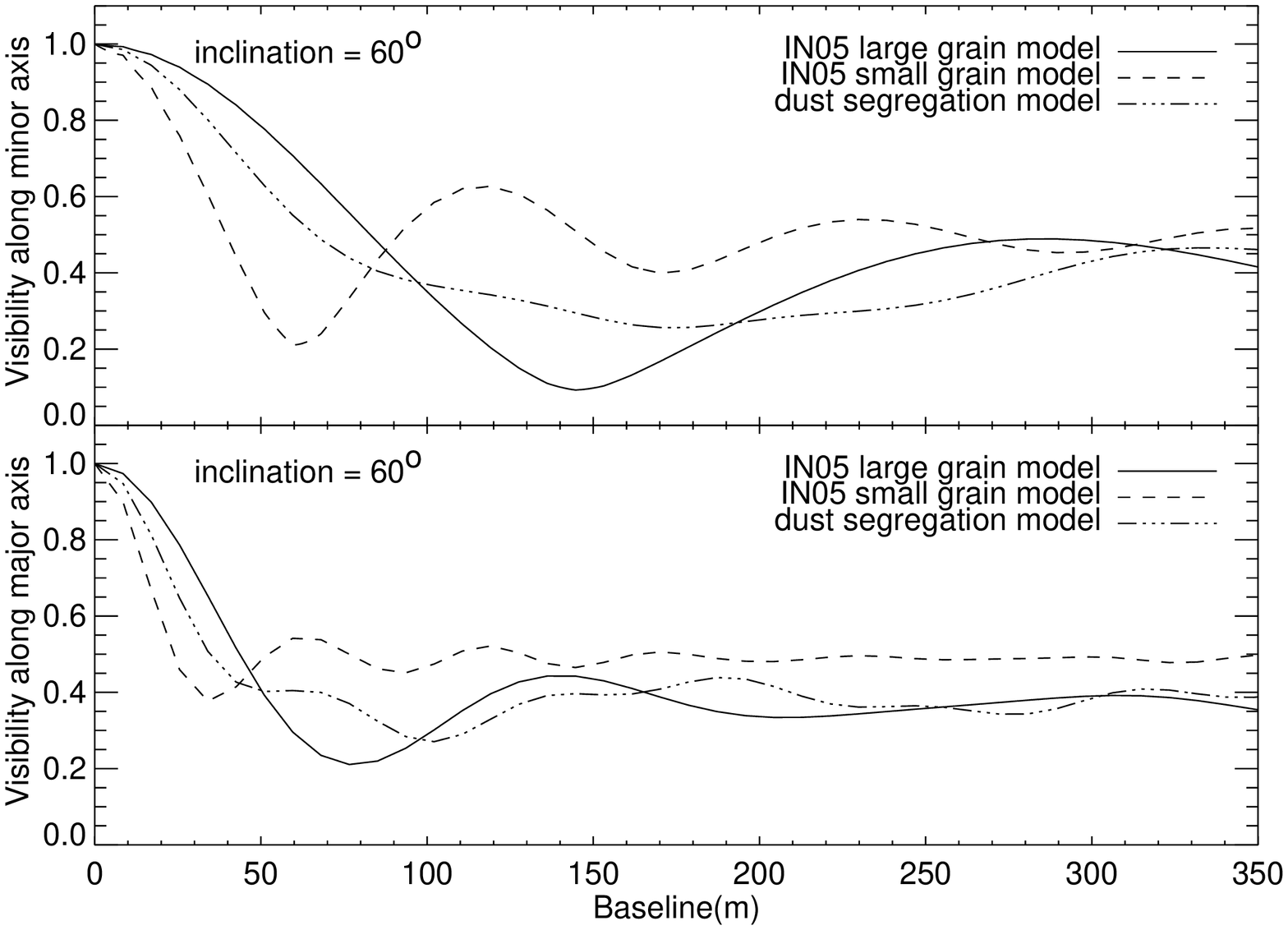}
\caption{2.2$\mu$m visibilities for the IN05 and dust segregation models. The panel
on the left shows the visibilities for a face-on disk and the right
panel shows visibilities computed along the major and minor axis for an
inclined disk.
\label{vis_2.2}}
\end{center}
\end{figure}

In order to illustrate the differences between the IN05 and dust segregation
models  we have computed synthetic images and visibilities for the
different models. At 2.2$\mu$m, all the emission comes from the rim
and the central star. The panels in Figure~\ref{dustsegim_2.2} show 2.2$\mu$m rim surface brightness for two inclination angles.  As
discussed previously, and is evident in Figure~\ref{dustsegim_2.2}, the
dust segregation rim is thicker than the IN05 rims. The dust
segregation models also distinguish themselves observationally from the
IN05 rims by their visibility curves. Figure~\ref{vis_2.2} shows plots
of 2.2$\mu$m visibilities calculated along the disk major and minor
axes. The visibilities at long baselines are more oscillatory in the
IN05 models. The IN05 rims are also more skewed at large inclination
angles which would give them a stronger closure phase signal
\citep{monnier2006}.

 At 10.7$\mu$m, emission comes from two distinct regions of the 
 circumstellar disk  as seen in Figure~\ref{dustsegim_10.7}. The rim
 has a high surface brightness at 10.7$\mu$m, but only contributes
  about 30-50\% of the integrated light (more precisely 38\% in the large grain
 model, 55\% in the dust segregation model and 32\% in the small grain
 model). In contrast to 2.2$\mu$m emission, a large fraction of the emission at 10.7$\mu$m comes from the flared disk
 \citep{keny1987, cg97} region on larger scales. Between the inner rim
 and the flared disk, there is a `shadow region' \citep{dullemond2001}
 which is not illuminated by the star directly. As discussed in
 \citet{boekel2005b}, the two distinct regions of emission cause a
 knee in the visibility curves (Figure~\ref{vis_10.7}). The size and
 shape of the knee is affected both by the inner rim as well as the
 flared disk structure \citep{boekel2005b}.

\citet{boekel2004a} studied the silicate feature in Herbig disks using
mid-infrared interferometry (with VLTI-MIDI), and showed that dust on AU
scales in Herbig disks has undergone significantly more thermal
processing as compared to the dust at tens of AU scale. The
discussion in this section demonstrates that the present generation of
long-baseline near- and mid-infrared interferometers can be used to
distinguish between inner-rim models, thus providing information on
dust processes that take place on even smaller scales (sub-AU).

\section{Discussion}
The DDN01 model was introduced to explain the near-IR excess observed in Herbig Ae systems. The near-IR excess in this model comes from a vertical inner dust rim which is frontally illuminated by the central star. As discussed in section \ref{intro}, the emission from a vertical rim geometry becomes highly skewed even at moderate line of sight inclination angles of the disk. Recent closure phase results on YSOs by \citet{monnier2006} do not find any evidence for the large skewness in near-IR emission predicted  by the DDN01 model. The \citet{monnier2006} results favor the existence of curved inner dust rims and discount the vertical inner wall of DDN01. The first and only detailed calculation existing in current literature for rim curvature was performed by IN05, where the rim gets curved because of a density dependence of dust sublimation temperatures. Initial fits to data \citep{isella2006} seem to suggest that the curved rim-model provides a better fit to the visibilities and near-IR SEDs of a few Herbig Ae systems, although detailed comparisons have not been made yet.

In this paper we introduced a new mechanism based on dust growth and settling to curve the inner rim. As a consequence of growth and settling, opacity in the rim mid-plane will be dominated by large grains whereas small grains dominate the opacity higher up in the disk photosphere. Large grains can exist much closer to the star than small grains since they cool more efficiently \citep[grain settling in the inner rim with large grains dominating the mid-plane opacity may be needed to explain near-IR SED and interferometry data on VVSer;][]{ponto2006}. A variation in grain size with disk scale height in the inner rim causes the rim to curve as shown in Figure~\ref{comparisona}. This curvature is on a much larger scale than the curvature produced in the IN05 models. Figures~\ref{vis_2.2} and \ref{vis_10.7} show that the two models can be distinguished based on their visibility curves.             

In the dust segregation model that we studied in section~\ref{model1}, the effects of dust growth and sedimentation on rim structure were calculated assuming that the dust sublimation temperature is independent of density. But in reality, both dust evolution and the density dependence of dust evaporation temperatures (INO5 model) probably simultaneously play a role in determining rim structure. We have shown that in the absence of strong turbulence the rim shape is predominantly determined by grain growth and settling. Adding a density dependence to dust sublimation temperatures will lower rim height (see Fig. 2) and extend it further into the disk by just a few tenths of an AU: a small perturbation compared to the effect of dust segregation on rim shape.

Calculations in this paper do not take into account effects of gas on dust rim geometry. Gas can affect the structure and location of the rim through accretion heating and by providing a source of opacity to stellar photons within the dust destruction radius. \citet{akeson2005a} have shown that gas accretion heating might contribute significantly to the near-IR excess observed in some TTauri stars and  play an important role in fixing the location of the dust destruction radius in these systems. \citet{eisner2006} argue that emission from gas within the inner rim is needed to explain spectrally dispersed near-IR interferometry data on Herbig Ae/Be systems.  The effects of gas on  dust rim structure and observed near-IR excesses in YSOs is not well understood.  Some theoretical efforts \citep[see for e.g. ][]{muzerolle2004} have been made in this direction but further progress would require detailed modeling of the relevant gas physics.

The dust rim in YSOs is a unique test-bed for theories of grain evolution. A number of observational results \citep{alessio2001, alessio2006, chiang2001, rettig2006} have found evidence for dust growth and settling in the outer regions (tens of AU scale) of YSO disks. In this work we show that long baseline near \& mid-IR interferometry provide an opportunity for similar studies on the inner rim at sub-AU scale. Furthermore, if the gas within the inner rim were to emit significantly in the near-IR (gas emission is not well constrained and preliminary results from CHARA on AB Aur suggest that this might be important) , then the baselines of CHARA($\sim$ 300m) are capable of partially resolving the hot gas emission, at-least for a few Herbig systems. Combined with accretion line diagnostics \citep{muzerolle2004}, near-IR interferometry could help pin down the structure of the inner gas disk.

In the future, we will compare SED and interferometry data with our model predictions for the inner rim. This will constrain dust growth, settling and turbulent mixing parameters that recent theoretical simulations \citep{Johan2005, fromang2006, turner2006} have begun to predict. These simulations show that dust grains up to 100 microns in size are well mixed with gas within one scale height. In the extreme case of gas and dust being well mixed at all scale heights, the rim shape will be controlled entirely by the density dependence of dust evaporation temperatures. Long-baseline near-IR interferometry can distinguish between various dust mixing scenarios and  furthermore, it  might even be possible to observe the time evolution of grain sedimentation processes by examining the shape of the inner rim as a function of stellar age. Detailed comparisons between model predictions and data will also provide insight into the role played by gas.

\section{Summary and Conclusions}
  
We have described a Monte Carlo approach to calculating
the structure of the inner rim in protoplanetary disks. The obvious
advantage of this method over an analytic approach is that it allows for a self-consistent
determination of the entire disk structure. Any effects of spatially
varying dust properties on disk structure (and the resulting effects
on SEDs and visibilities) can also be easily studied.

\begin{figure}[thb]
\begin{center}
\includegraphics[angle=90,height=4.55in, width=7.0in]{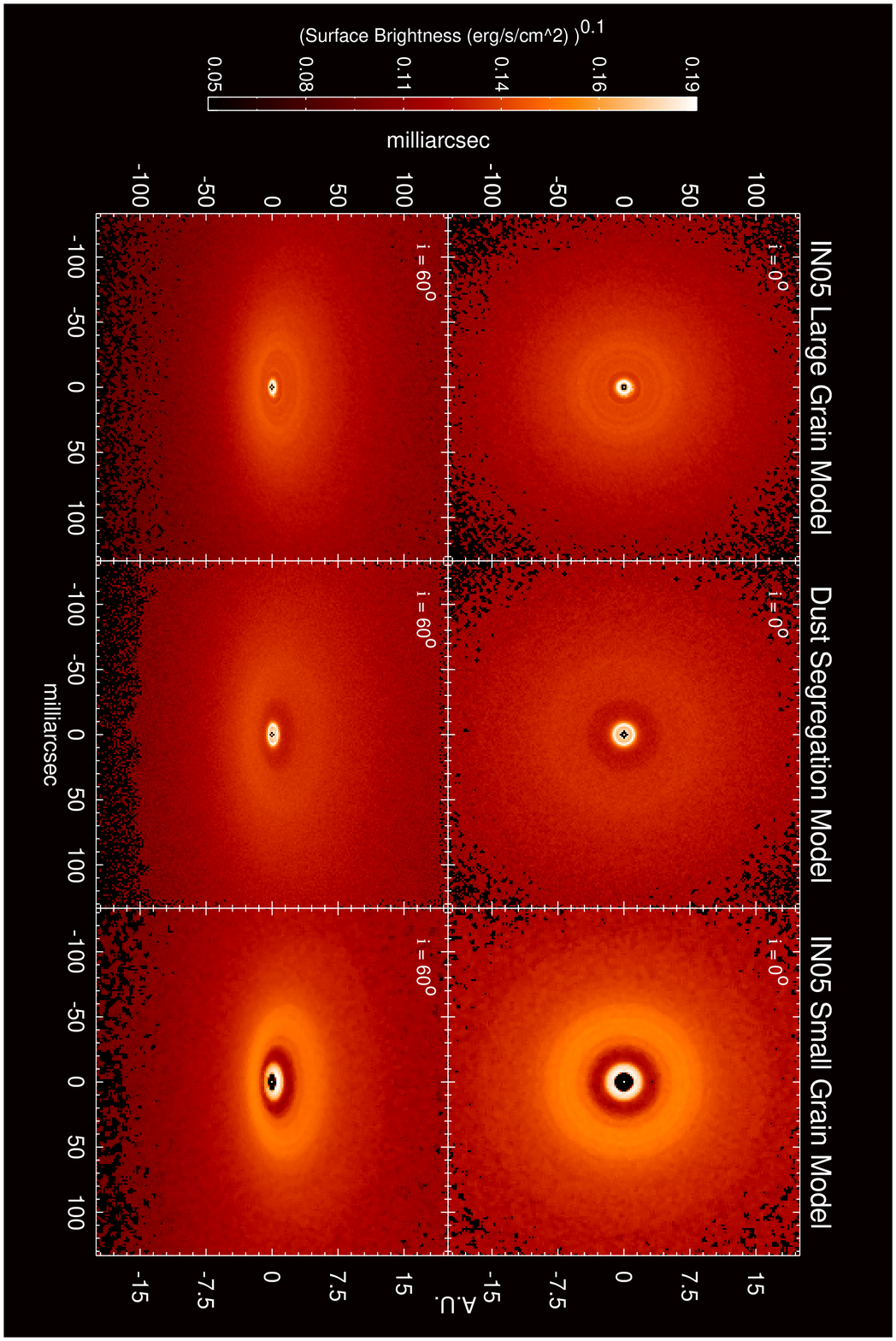}
\caption{Synthetic 10.7$\mu$m images for the different
rim models discussed in the text. The panels on the left and right are
IN05 rims computed for 1.2$\mu$m (large grain) and 0.1$\mu$m (small grain)
silicate dust. The center panels are images for the dust
segregation model. The star is placed at 150pc with the stellar
parameters described in Table~\ref{table1}. The star is unresolved
at the image scale and is just one bright pixel at the center.
\label{dustsegim_10.7}}
\end{center}
\end{figure}

\begin{figure}[thb]
\begin{center}
\includegraphics[angle=90,width=3in]{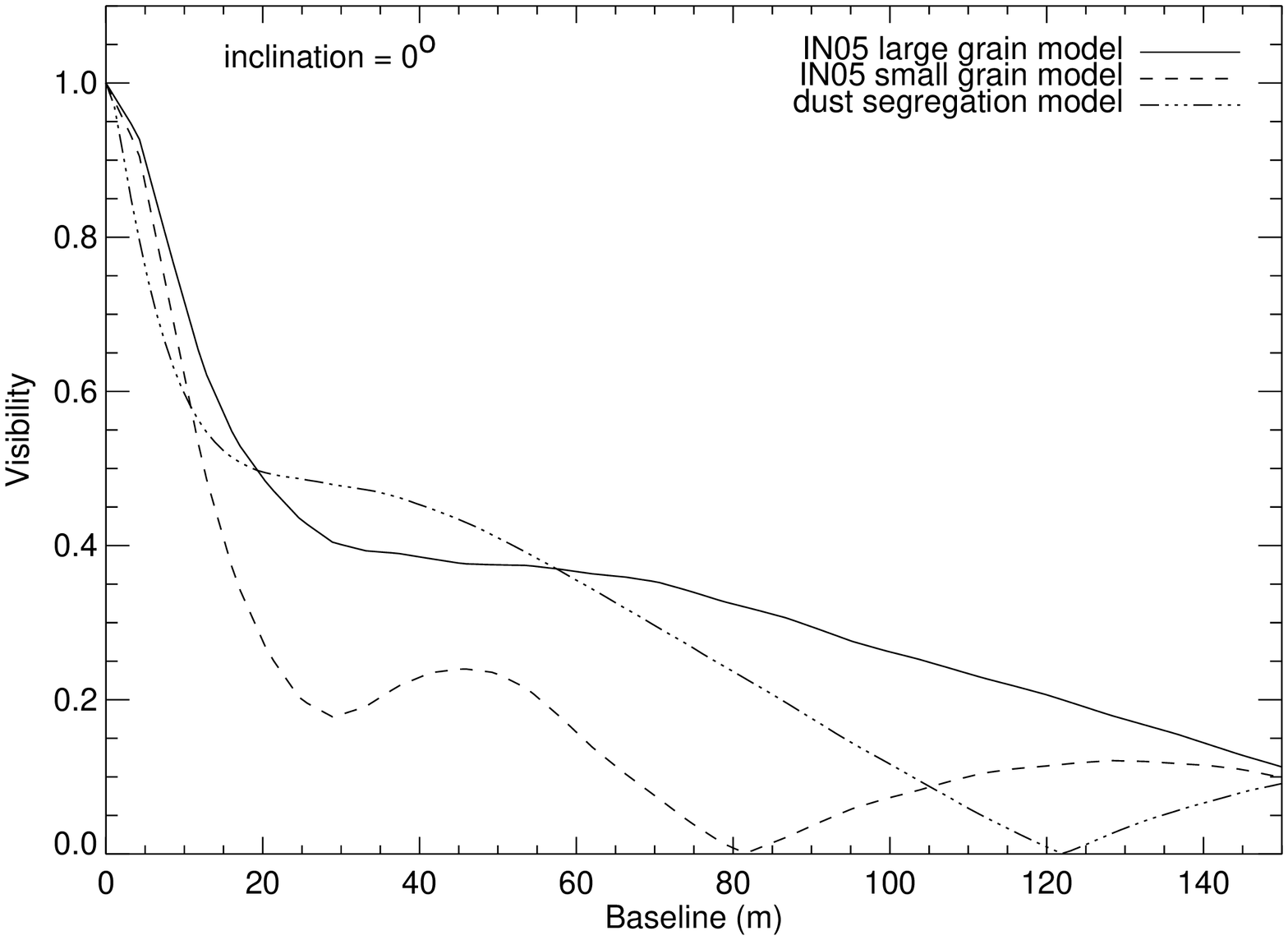}
\hphantom{.....}
\includegraphics[angle=90,width=3in]{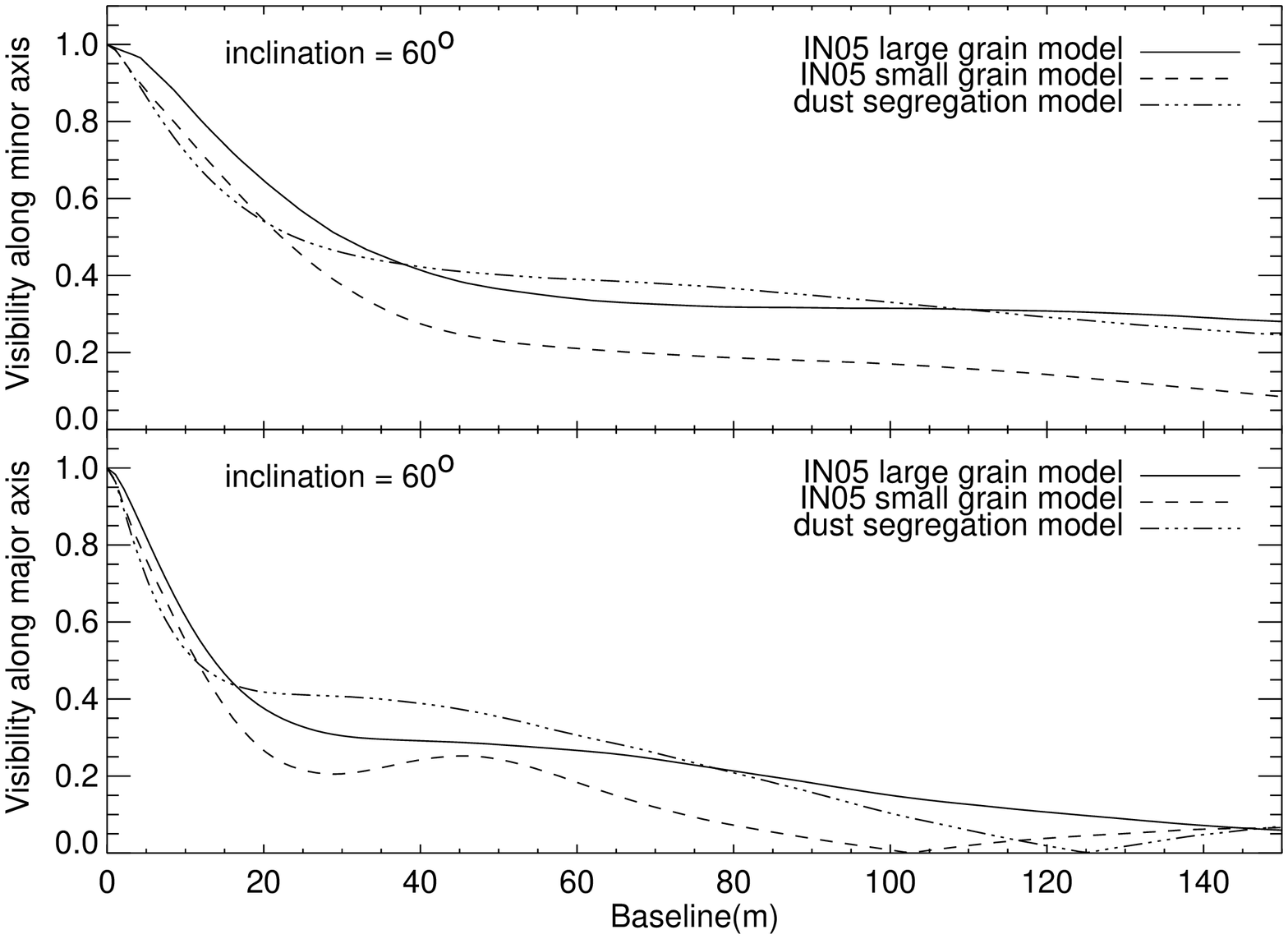}
\caption{10.7$\mu$m visibilities for the IN05 and dust segregation models. The
panel on the left shows the visibilities for a face-on disk and the
right panel shows visibilities computed along the major and minor axis
for an inclined disk.
\label{vis_10.7}}
\end{center}
\end{figure}

We have shown that dust growth and sedimentation may play a dominant role
in determining the structure of the inner rim. The difference in
cooling properties between grains of different sizes, combined with
sublimation and a vertical gradient in dust sizes in the disk, leads
to inner rims that are extremely curved (over scales $>$ 0.7 AU in cylindrical radius for an A0 star, see Figure \ref{comparisona}). Synthetic images (Figures \ref{dustsegim_2.2} \& \ref{dustsegim_10.7}) and visibilities (Figures \ref{vis_2.2} \& \ref{vis_10.7}) for the rim models have been computed, highlighting
differences between various theoretical models. 

The inner rim dominates near-IR emission from circumstellar
disks. Studies in the near-IR thus isolate the rim from rest of the disk
and are therefore extremely powerful tools for characterizing rim
properties. In the last decade or so, near-IR interferometers
\citep{rmg1999, rmg2001, eisner2003, eisner2004, akeson2005a,
akeson2005b} have been used to determine the sizes of inner rims in a
number of Herbig Ae/Be and TTauri stars. This has helped in
establishing a tight correlation between the radii of the rims and the
luminosities of the central stars \citep{monnier2002a, monnier2005}. The
current generation of near-IR interferometers (CHARA, VLTI) are
capable of taking the next natural step---determining the geometry of
the inner rims. As shown in this paper and IN05, the geometry of the
inner rim is tied to multiple dust properties such as grain evolution,
grain sedimentation (which depends on the strength of turbulence), and
dust sublimation temperatures (IN05). Once the inner rim shape is
constrained, mid-IR interferometry (as seen in
Figure~\ref{dustsegim_10.7}, both the rim and the flared disk region
emit in the mid IR) can  be used to assess the geometry
of the outer, flared-disk component.
  
A significant number of Herbig Ae/Be stars have multi-wavelength
(near-IR, mid-IR and mm-interferometry, and spectroscopy over a large
wavelength range) data available. In the near future  we will try to
generate disk models that simultaneously fit interferometry and
spectral data over multiple wavelengths for a small sample of
stars. This will be a stringent test for current disk models and
should help alleviate model degeneracy that plagues single-wavelength
studies.

\acknowledgements
AT wishes to thank R. Kurosawa for
parallelizing the image making modules in {\sc torus}, A. Isella for
providing data points for the IN05 curves in Figure~\ref{comparisona}
and N. Calvet for useful discussions. This project was partially
supported by NASA grant 050283. This publication makes use of NASA's
Astrophysics Data System Abstract Service.

\appendix

\section{Appendix}
Here, we derive a general analytic formula for the rim shape of a dust
segregation model (a case with specific numerical values for
parameters, was discussed in section \S\ref{analytics}).

Let the densities of the two dust components in the circumstellar disk be given by 
\begin{equation}
\rho_{\rm large} = A_{\rm large}\rho _{o\_{\rm gas}}\exp \left[ -\frac{1}{2}\frac{z^2}{(\zeta_{\rm large} h)^2}\right]
\end{equation}

\begin{equation}
\rho_{\rm small} = A_{\rm small}\rho_{o\_{\rm gas}} \exp \left[ -\frac{1}{2}\frac{z^2}{(\zeta_{\rm small} h)^2}\right]  
\end{equation}

\noindent where $\rho_{\rm large}$ and $\rho_{\rm small}$ are the densities of large and small grains respectively and $\rho_{o\_{\rm gas}}$ is the gas density in the mid-plane. $A_{\rm large}$ and $A_{\rm small}$ are normalization factors
for the dust densities, $z$ is the height from the disk mid-plane, $h$
is the gas scale height and $\zeta_{\rm large}$ and $\zeta_{\rm small}$ are the scale
heights for large and small grains as fractions of the gas scale
height.

The ratio of Planck mean opacities at the dust evaporation temperature
and the stellar effective temperature - $\epsilon (z)$, then becomes

\begin{equation}
\epsilon\left (z\right) = \left[ \frac {A_{\rm large}K_{\rm
    P_{\rm large}}(T_{\rm evp}) + A_{\rm small}K_{\rm P_{\rm small}}(T_{\rm evp}) \exp
    \left(\frac{1}{2}\frac{z^2(\zeta^2_{\rm small}-\zeta^2_{\rm large})}{h^2\zeta^2_{\rm
    large}\zeta^2_{\rm small}}\right)}{A_{\rm large}K_{\rm
    P_{\rm large}}(T_*) + A_{\rm small}K_{\rm P_{\rm small}}(T_*) \exp \left(\frac{1}{2}\frac{z^2(\zeta^2_{\rm small}-\zeta^2_{\rm large})}{h^2\zeta^2_{\rm large}\zeta^2_{\rm small}}\right)}\right]
\end{equation}

\noindent where $K_{\rm P_{\rm large}}$ and $K_{\rm P_{\rm small}}$ are Planck mean opacities for large
and small grains at the specified temperatures, $T_{\rm evp}$ is the dust
evaporation temperature and $T_*$ is the stellar effective
temperature.

Neglecting local accretion heating, the dust destruction radius $R_{\rm evp}$ \citep[see][]{isella2006,
calvet1991} is

\begin{equation}
R_{\rm evp}\left[\mbox{AU}\right] =  0.034\left[\mbox{AU}\right]\left(\frac{1500\K}{T_{\rm evp}}\right)^2\left[{ \left(\frac{L_*}{L_{\odot}}\right)\left(B + \frac{1}{\epsilon(z)}\right) }\right]^{1/2},
\end{equation}
$L_*$ is luminosity of the star in solar units. B is a dimensionless  diffuse heating term, characterizing the rim optical depth to its own thermal emission.

The gas scale height $h$ at $R_{evp}$ is estimated to be \citep[see eqs. {[7]} and {[8]} from][]{cg97}
\begin{equation}
 h\left[AU\right] =  R_{\rm evp}\left[\mbox{AU}\right] \left[{\left(\frac{T_{\rm gas}}{4\times 10^{7}\K}\right)\left(\frac{M_{\odot}}{M_*}\right) \left(\frac{R_{\rm evp}}{4.64\times 10^{-3}\mbox{ AU}}\right)}\right]^{1/2}
\end{equation}
where $T_{\rm gas}$ is the temperature of the bulk of the gas (assumed to be isothermal) and $M_*$ is mass of the
star in solar masses.

Combining equations (A3), (A4), and (A5), the height of the evaporation front as a
function of distance along the disk mid-plane from the star becomes
\begin{eqnarray}
 z[\mbox{AU}] =  R_{\rm evp}[\mbox{AU}]\left[2\zeta^2_{\rm diff}{\mbox{ln}\left(\frac{A_{\rm large}K_{\rm
 P_{\rm large}}(T_*)-Y A_{\rm large}K_{\rm P_{\rm large}}(T_{\rm evp})}{ YA_{\rm small}K_{\rm P_{\rm small}}(T_{\rm evp}) -A_{\rm small}K_{\rm
 P_{\rm small}}(T_*)}\right)}\right]^{1/2}  \nonumber \\
\times \left[{\left(\frac{T_{\rm gas}}{4\times 10^{7}\K}\right)\left(\frac{M_{\odot}}{M_*}\right)\left(\frac{R_{\rm evp}}{4.64\times 10^{-3}\mbox{ AU}}\right)}\right]^{1/2}
\end{eqnarray}
 where $Y$ and $\zeta_{\rm diff}$  are dimensionless terms given by 
\begin{equation}
Y =  \left(\frac{ 1.71\times 10^{-10}\mbox{AU}^{-2} \K^{-4}R^2_{evp}T^4_{\rm evp}}{L_*/L_{\odot}} - B\right)
\end{equation}

\begin{equation}
\zeta^2_{\rm diff} =  \frac{\zeta^2_{\rm small}\zeta^2_{\rm large}}{\zeta^2_{\rm small}-\zeta^2_{\rm large}}
\end{equation}

\bibliographystyle{apj}
\bibliography{model_temp}

\clearpage

\begin{deluxetable}{rclrrrcl} 
\tablecolumns{8} 
\tablewidth{0pc} 
\tablecaption{Basic properties of central star and the circumstellar disk\label{table1}} 
\tablehead{ 
\colhead{}    &  \multicolumn{2}{c}{Star} &   \colhead{}   & \colhead{} & \colhead{} &
\multicolumn{2}{c}{ Circumstellar Disk} }
\startdata 

  & Mass  &  2.5$\mbox{ M}_{\odot}$ & &  &  & 
 Surface Density & $\Sigma (r) =2000(r/\mbox{AU})^{-1.5} \mbox{ g }\mbox{cm}^{-2}$\\  

 & $\mbox{T}_{\mbox{eff}}$ & $10,000\mbox{ K}$ &  &  &  &
Disk outer-Radius &200 AU\\

 &Luminosity &47$\mbox{ L}_{\odot}$ & & & &
  Mass &3.8 $\times$ $10^{-2}$$\mbox{ M}_{\odot}$\\  
 &Distance&150 pc & & & & \\
\enddata
\end{deluxetable}

\clearpage

%
%



%









\clearpage
\end{document}